\documentclass[10pt]{article}
\usepackage{axodraw}
\usepackage{epsf}

\newcommand{\be}{\begin{equation}}
\newcommand{\ee}{\end{equation}}
\newcommand{\ba}{\begin{eqnarray}}
\newcommand{\ea}{\end{eqnarray}}

\newcommand{\eq}{Eq.~}
\newcommand{\nr}[1]{(\ref{#1})}
\newcommand{\nn}{\nonumber \\}
\newcommand{\fr}[2]{{\frac{#1}{#2}\,}}
\renewcommand{\(}{\left(}
\renewcommand{\)}{\right)}

\newcommand{\lk}{\left[}
\newcommand{\rk}{\right]}

\newcommand{\e}{\epsilon}

\newcommand{\ep}{\;\e}
\newcommand{\eqs}{Eqs.~}
\newcommand{\order}[1]{{\cal O}\(#1\)\vphantom{\fr12}}

\def\Asc(#1,#2)(#3,#4,#5){\CArc(#1,#2)(#3,#4,#5)}
\def\Lsc(#1,#2)(#3,#4){\Line(#1,#2)(#3,#4)}
\def\Ahh(#1,#2)(#3,#4,#5){\DashCArc(#1,#2)(#3,#4,#5){1}}
\def\Lhh(#1,#2)(#3,#4){\DashLine(#1,#2)(#3,#4){1}}
\def\Aqu(#1,#2)(#3,#4,#5){\ArrowArc(#1,#2)(#3,#4,#5)}
\def\Aaqu(#1,#2)(#3,#4,#5){\ArrowArcn(#1,#2)(#3,#5,#4)}
\def\scfc{0.7}  
\def\phgt{21}   
\def\pwc{21}    
\def\pwcb{31.5} 
\newcommand{\PIC}[4]{\;\parbox[c]{#2 pt}{\begin{picture}(#2,#3)(0,0)
\SetWidth{1.0}\SetScale{#4} #1 \end{picture}}\;}
\newcommand{\pic}[1]{\PIC{#1}{\pwc}{\phgt}{\scfc}}
\newcommand{\picb}[1]{\PIC{#1}{\pwcb}{\phgt}{\scfc}}
\def\TopoVR(#1){\pic{#1(15,15)(15,-90,270)}}
\def\ToptVS(#1,#2,#3){\pic{#1(15,15)(15,0,180) #2(15,15)(15,180,360)%
 #3(30,15)(0,15)}}
\def\ToprVM(#1,#2,#3,#4,#5,#6){\pic{#3(15,15)(15,-30,90) #1(15,15)(15,90,210)%
 #2(15,15)(15,210,330) #5(2,7.5)(15,15) #6(15,15)(15,30) #4(28,7.5)(15,15)}}
\def\ToprVV(#1,#2,#3,#4,#5){\!\!\picb{#2(26.25,15)(15,256,76)%
 #3(30,30)(15,30) #1(18.75,15)(15,104,284) #4(15,30)(22.5,0)%
 #5(30,30)(22.5,0)}\!\!}
\def\ToprVB(#1,#2,#3,#4){\picb{#1(30,15)(15,-120,120) #2(30,15)(15,120,240)%
 #3(15,15)(15,60,300) #4(15,15)(15,-60,60)}}
\def\TopfVX(#1,#2,#3,#4,#5,#6,#7,#8,#9){\picb{#1(15,15)(15,90,270)%
 #2(30,15)(15,-90,90) #4(30,30)(15,30) #3(15,0)(30,0) #6(15,0)(15,15)%
 #5(15,15)(30,30) #8(15,30)(20,25) #8(25,20)(30,15) #7(30,15)(30,0)%
 #9(15,15)(30,15)}}
\def\TopfVH(#1,#2,#3,#4,#5,#6,#7,#8,#9){\picb{#1(15,15)(15,90,270)%
 #2(30,15)(15,-90,90) #4(30,30)(15,30) #3(15,0)(30,0) #6(15,0)(15,15)%
 #5(15,15)(15,30) #8(30,30)(30,15) #7(30,15)(30,0) #9(15,15)(30,15)}}
\def\TopfVW(#1,#2,#3,#4,#5,#6,#7,#8){\pic{#1(15,15)(15,90,180)%
 #3(15,15)(15,180,270) #2(15,15)(15,270,360) #4(15,15)(15,0,90)%
 #5(15,15)(15,30) #7(15,15)(15,0) #6(0,15)(15,15) #8(30,15)(15,15)}}
\def\TopfVWdot(#1,#2,#3,#4,#5,#6,#7,#8){\pic{#1(15,15)(15,90,180)%
 #3(15,15)(15,180,270) #2(15,15)(15,270,360) #4(15,15)(15,0,90)%
 #5(15,15)(15,30) #7(15,15)(15,0) #6(0,15)(15,15) #8(30,15)(15,15)%
 \Vertex(22,15){2}}}
\def\TopfVWlap(#1,#2,#3,#4,#5,#6,#7,#8){\pic{#1(15,15)(15,90,180)%
 #3(15,15)(15,180,270) #2(15,15)(15,270,360) #4(15,15)(15,0,90)%
 #5(15,15)(15,30) #7(15,15)(15,0) #6(0,15)(15,15) #8(30,15)(15,15)%
 \Text(2,19)[br]{$\scriptstyle 1$}\Text(20,2)[tl]{$\scriptstyle 2$}}}
\def\TopfVV(#1,#2,#3,#4,#5,#6,#7,#8){\!\!\picb{#2(26.25,15)(15,256,346)%
 #3(26.25,15)(15,-14,76) #4(30,30)(15,30) #1(18.75,15)(15,104,284)%
 #7(22.5,0)(15,30) #6(30,30)(26.25,15) #8(26.25,15)(22.5,0)%
 #5(25.25,15)(39.8,11.4)}\!\!}
\def\TopfVB(#1,#2,#3,#4,#5,#6,#7){\picb{#2(30,15)(15,-120,120)%
 #6(30,15)(15,120,180) #5(30,15)(15,180,240) #1(15,15)(15,60,300)%
 #4(15,15)(15,-60,0) #3(15,15)(15,0,60) #7(30,15)(15,15)}}
\def\TopfVBdot(#1,#2,#3,#4,#5,#6,#7){\picb{#2(30,15)(15,-120,120)%
 #6(30,15)(15,120,180) #5(30,15)(15,180,240) #1(15,15)(15,60,300)%
 #4(15,15)(15,-60,0) #3(15,15)(15,0,60) #7(30,15)(15,15)%
 \Vertex(28,22){2}}}
\def\TopfVBlap(#1,#2,#3,#4,#5,#6,#7){\picb{#2(30,15)(15,-120,120)%
 #6(30,15)(15,120,180) #5(30,15)(15,180,240) #1(15,15)(15,60,300)%
 #4(15,15)(15,-60,0) #3(15,15)(15,0,60) #7(30,15)(15,15)%
 \Text(9,15)[r]{$\scriptstyle 1$}\Text(22.5,15)[l]{$\scriptstyle 2$}}}
\def\TopfVN(#1,#2,#3,#4,#5,#6,#7){\picb{#1(15,15)(15,90,270)%
 #2(30,15)(15,-90,90) #4(30,30)(15,30) #3(15,0)(30,0)%
 #5(15,0)(15,30) #6(30,30)(30,0) #7(15,30)(30,0)}}
\def\TopfVU(#1,#2,#3,#4,#5,#6,#7){\pic{#3(15,15)(15,0,90)%
 #2(15,15)(15,90,180) #4(15,15)(15,180,270) #1(15,15)(15,270,360)%
 #6(0,15)(15,30) #7(15,0)(0,15) #5(30,15)(15,0)}}
\def\TopfVT(#1,#2,#3,#4,#5,#6){\pic{#1(15,15)(15,90,210)%
 #2(15,15)(15,210,330) #3(15,15)(15,-30,90) #4(2,7.5)(15,30)%
 #6(28,7.5)(2,7.5) #5(15,30)(28,7.5)}}
\def\TopLV(#1,#2,#3,#4,#5,#6){\!\!\picb{#2(26.25,15)(15.5,256,76)%
 #3(30,30)(15,30) #1(18.75,15)(15.5,104,284) #4(15,30)(22.5,0)%
 #5(30,30)(22.5,0) #6(15,17.8)(19.3,292.8,39.1)}\!\!}
\def\TopfVBB(#1,#2,#3,#4,#5){\picb{#1(30,15)(15,-120,120)%
 #2(30,15)(15,120,240) #3(15,15)(15,60,300) #4(15,15)(15,-60,60)%
 #5(22.5,3)(22.5,27)}}
\def\TopfVBBdot(#1,#2,#3,#4,#5){\picb{#1(30,15)(15,-120,120)%
 #2(30,15)(15,120,240) #3(15,15)(15,60,300) #4(15,15)(15,-60,60)%
 #5(22.5,3)(22.5,27) \Vertex(22.5,10){2} \Vertex(22.5,20){2}}}
\def\TopfVBBlap(#1,#2,#3,#4,#5){\picb{#1(30,15)(15,-120,120)%
 #2(30,15)(15,120,240) #3(15,15)(15,60,300) #4(15,15)(15,-60,60)%
 #5(22.5,3)(22.5,27)%
 \Text(2,10.5)[l]{$\scriptstyle 1$}\Text(29.5,10.5)[r]{$\scriptstyle 2$}}}

\def\one{\TopoVR(\Asc)}
\def\two{\ToptVS(\Asc,\Asc,\Lsc)}
\def\threeM{\ToprVM(\Asc,\Asc,\Asc,\Lsc,\Lsc,\Lsc)}
\def\threeV{\ToprVV(\Asc,\Asc,\Lsc,\Lsc,\Lsc)}
\def\threeB{\ToprVB(\Asc,\Asc,\Asc,\Asc)}
\def\topoEX{\TopfVX(\Asc,\Asc,\Lsc,\Lsc,\Lsc,\Lsc,\Lsc,\Lsc,\Lsc)}
\def\topoI{\TopfVH(\Asc,\Asc,\Lsc,\Lsc,\Lsc,\Lsc,\Lsc,\Lsc,\Lsc)}
\def\topoII{\TopfVW(\Asc,\Asc,\Asc,\Asc,\Lsc,\Lsc,\Lsc,\Lsc)}
\def\topoIII{\TopfVV(\Asc,\Asc,\Asc,\Lsc,\Lsc,\Lsc,\Lsc,\Lsc)}
\def\topoIV{\TopfVB(\Asc,\Asc,\Asc,\Asc,\Asc,\Asc,\Lsc)}
\def\topoV{\TopfVN(\Asc,\Asc,\Lsc,\Lsc,\Lsc,\Lsc,\Lsc)}
\def\topoVI{\TopfVU(\Asc,\Asc,\Asc,\Asc,\Lsc,\Lsc,\Lsc)}
\def\topoVIII{\TopfVT(\Asc,\Asc,\Asc,\Lsc,\Lsc,\Lsc)}
\def\topoIX{\TopLV(\Asc,\Asc,\Lsc,\Lsc,\Lsc,\Asc)}
\def\topoXII{\TopfVBB(\Asc,\Asc,\Asc,\Asc,\Lsc)}

\def\topoIIxtra{\TopfVWdot(\Asc,\Asc,\Asc,\Asc,\Lsc,\Lsc,\Lsc,\Lsc)}
\def\topoIVxtra{\TopfVBdot(\Asc,\Asc,\Asc,\Asc,\Asc,\Asc,\Lsc)}
\def\topoXIIxtra{\TopfVBBdot(\Asc,\Asc,\Asc,\Asc,\Lsc)}

\catcode`@=11
\newcount\@tempcntc
\def\@citex[#1]#2{\if@filesw\immediate\write\@auxout{\string\citation{#2}}\fi
  \@tempcnta\z@\@tempcntb\m@ne\def\@citea{}\@cite{\@for\@citeb:=#2\do
    {\@ifundefined
       {b@\@citeb}{\@citeo\@tempcntb\m@ne\@citea\def\@citea{,}{\bf ?}\@warning
       {Citation `\@citeb' on page \thepage \space undefined}}%
    {\setbox\z@\hbox{\global\@tempcntc0\csname b@\@citeb\endcsname\relax}%
     \ifnum\@tempcntc=\z@ \@citeo\@tempcntb\m@ne
       \@citea\def\@citea{,}\hbox{\csname b@\@citeb\endcsname}%
     \else
      \advance\@tempcntb\@ne
      \ifnum\@tempcntb=\@tempcntc
      \else\advance\@tempcntb\m@ne\@citeo
      \@tempcnta\@tempcntc\@tempcntb\@tempcntc\fi\fi}}\@citeo}{#1}}
\def\@citeo{\ifnum\@tempcnta>\@tempcntb\else\@citea\def\@citea{,}%
  \ifnum\@tempcnta=\@tempcntb\the\@tempcnta\else
   {\advance\@tempcnta\@ne\ifnum\@tempcnta=\@tempcntb \else \def\@citea{--}\fi
    \advance\@tempcnta\m@ne\the\@tempcnta\@citea\the\@tempcntb}\fi\fi}
\catcode`@=12

\date{}

\begin{document}


\title{
\centerline{\normalsize \mbox{} \hfill HIP-2003-55/TH}
\vskip-1ex
\centerline{\normalsize \mbox{} \hfill MIT-CTP 3431} 
\vskip-1ex
\centerline{\normalsize \mbox{} \hfill hep-ph/0311323} 
\vskip2ex
High-precision evaluation of four-loop vacuum bubbles\\ in three dimensions}
\author{\small Y.~Schr\"oder$^a$, A.~Vuorinen$^b$\\
{\small\it $^a$ Center for Theoretical Physics, MIT, Cambridge, MA 02139, 
USA}\\
{\small\it $^b$ Department of Physical Sciences and Helsinki Institute of Physics}\\
{\small\it P.O Box 64, FIN-00014 University of Helsinki, Finland}}

\maketitle

\begin{abstract}
In this letter we present a high-precision evaluation of the expansions
in $\e=(3-d)/2$ of (up to) four-loop scalar vacuum master integrals,
using the method of difference equations developed by Laporta.
We cover the complete set of fully massive master integrals.

\medskip

\end{abstract}



\section{Introduction}

Higher-order perturbative computations have become a necessity
in many areas of theoretical physics, be it
for high-precision tests of QED, QCD and the standard model, 
or for studying critical phenomena in condensed matter systems.

Most recent investigations employ a highly automated approach,
utilizing algorithms that can be implemented on computer
algebra systems, in order to handle the growing numbers
of diagrams as well as integrals which occur at higher loop
orders.

Computations can be divided into four key steps.
First, the complete set of diagrams including symmetry factors
has to be generated. 
For a detailed description of an algorithm for this step 
for the case of vacuum topologies, see \cite{Kajantie:2001hv}.
Second, after specifying the Feynman rules, the color- and 
Lorentz-algebra has to be worked out. 
Third, within dimensional regularization,
massive use of the integration-by-parts (IBP) technique \cite{Chetyrkin:qh}
to derive linear relations between different Feynman integrals 
in conjunction with an ordering prescription can be used 
to reduce the (typically large number of) integrals to a basis
of (typically a few) master integrals \cite{Laporta:2001dd}.  
Practical notes as well as a classification of vacuum master
integrals is given in \cite{Schroder:2002re}.
Fourth, the master integrals have to be solved, either fully
analytically, or in an expansion around the space-time dimension $d$
of interest. 

It is the fourth step that we wish to address here. 
While most work has been and is being devoted to $d=4$, 
perturbative results in lower dimensions are needed
for applications in condensed matter systems, as well
as in the framework of dimensionally reduced effective 
field theories for thermal QCD, where recent efforts have 
made four-loop contributions an issue \cite{Kajantie:2002wa}. 

A very important subset of master integrals are fully massive 
vacuum (bubble) integrals, since they constitute a main building
block in asymptotic expansions (see e.g. \cite{Misiak:1994zw}). 
They are also useful for massless
theories, when a propagator mass is introduced as an intermediate 
infrared regulator \cite{Schroder:2003uw}.

The main purpose of this note is to numerically 
compute the complete set of fully massive vacuum 
master integrals in terms of a high-precision $\e$\/-expansion
in $d=3-2\e$ dimensions, in complete analogy with the four-dimensional
work of S.~Laporta \cite{Laporta:2002pg}. 

The plan of the paper is as follows.
In Section \ref{se:two}, we give a brief review of the method 
of difference equations applied to vacuum integrals.
In Section \ref{se:three}, we discuss the actual implementation
of the algorithm.
In Section \ref{se:four}, we display our numerical results for
the truncated power series expansions in $\e$ of
all fully massive master integrals, up to four-loop level,
in $d=3-2\e$.


\section{\label{se:two}
The evaluation of master integrals through difference equations}

The method we have chosen to compute the coefficients of the
truncated power series expansions of the master integrals 
is based on constructing difference equations for the integrals 
and then solving them numerically using factorial series. 
This approach was recently developed in 
Ref.~\cite{Laporta:2001dd}, and below we briefly
summarize its basic concepts following the notation of the original paper, 
which contains a much more detailed presentation on the subject.
While the method is completely general  
as it applies to arbitrary kinematics, masses and topologies 
\cite{Laporta:2001rc}, 
our brief summary is somewhat adapted to the specific case 
of massive vacuum integrals.

The main idea is to attach an arbitrary power $x$ to one of the
lines of a master integral $U$,
\ba
U(x)&\equiv&\int\fr{1}{D_1^xD_2...D_{N}},
\ea
where the $D_i=(p_i^2+1)$ denote inverse scalar propagators.
In our case all of these share the same mass $m$, which we have 
therefore set to $1$, noting that it can be restored in the
end as a trivial dimensional prefactor of each integral.
The original integral is then just $U=U(1)$.
Depending on the symmetry properties of the integral,
there can be different choices for the `special' line with
the arbitrary power $x$, but in the limit $x=1$ they all reduce to 
the original integral $U$. 
This degeneracy can (and will later) be used for non-trivial 
checks of the method.

Employing IBP identities in a systematic way, it is possible
to derive a linear difference equation obeyed by the generalized
master integral $U(x)$,
\ba \label{diffeqn}
\sum_{j=0}^{R} p_{j}(x)U(x+j)&=&F(x), \label{pushdown}
\ea
where $R$ is a finite
positive integer and the coefficients $p_{j}$ are polynomials in $x$
(and the space-time dimension $d$). The function $F$ on the 
r.h.s. is a linear combination of
functions analogous to $U(x)$ but derived from simpler master integrals, 
i.e. integrals containing a smaller number of loops and/or propagators. 

The general solution of this kind of an equation is the sum of a special
solution of the full equation, $U_0(x)$, and all solutions of the homogeneous equation ($F=0$),
\ba \label{gensoln}
U(x)&=&U_0(x)+\sum_{j=1}^R U_j(x),
\ea
where each ($j=0,...,R$)
\ba \label{soln}
U_j(x)=\mu_j^x \sum_{s=0}^\infty a_j(s) \fr{\Gamma(x+1)}{\Gamma(x+1+s-K_j)}
\ea
is a factorial series\footnote{For a rigorous definition of the concept as 
well as a motivation for this kind of an ansatz, we refer the reader to Ref. \cite{Laporta:2001dd}.}. 
Substituting into \eq\nr{diffeqn}, 
one obtains the coefficients $\mu$ and $K$ 
(the latter being a function of $d$), as well as recursion relations 
for the $x$\/-independent coefficients $a(s)$ (being functions of $d$ 
as well) for each solution. 
For the homogeneous solutions, these recursion relations relate all 
coefficients to their value at $s=0$, $a_j(s)=c_j(s)\,a_j(0)$, 
where the $c_j(s)$ are rational functions (of $d$ as well).
For the special solution, the $a_0(s)$ are completely fixed in terms of
the inhomogeneous part $F(x)$,  
consisting of `simpler' integrals which are assumed to
already be known in terms of their factorial series expansions.

What remains to be done is to fix the $x$\/- and $s$\/-independent
constants $a_j(0)$, $j\neq 0$, in order to determine the weights of the
different homogeneous solutions. 
To this end, it is most useful to study the behavior of $U(x)$ 
at large $x$, where the first factor in
\ba
U(x)&=&\int\fr1{(p_1^2+1)^x}\, g(p_1)
\ea
peaks strongly around $p_1^2=0$. 
Hence, the large-$x$ behavior of the modified master integral
is determined by the small-momentum expansion of the two-point
function $g(p_1)$, which has one loop less than the original 
vacuum integral. 
In fact, for all cases we cover here, the first coefficient
in the asymptotic expansion suffices. This is furthermore particularly
simple, since it factorizes into
a one-loop bubble carrying the large power $x$ and a lower-loop
vacuum bubble $g(0)$, which corresponds to $U(x)$ with its `special'
line cut away,
\ba
\lim_{x\rightarrow\infty} U(x) &=& \lk \int\fr1{(p_1^2+1)^x}\rk 
\times \lk \vphantom{\int} g(0) \rk
\;\sim\; (1)^x x^{-d/2} g(0) \;.
\ea
A comparison with the large-$x$ behavior of 
\eqs\nr{gensoln}, \nr{soln}, proportional to $\sum_j \mu_j^x a_j(0) x^{K_j}$,
can now be used to fix the $a_j(0)$, 
of which maximally one will
turn out to be non-zero for our set of integrals.

Having the full solution at hand, we have in principle completed our 
entire task, as in the limit $x=1$ we recover from $U(x)$ the value
of the initial integral. 
Let us, however, add a couple of practical remarks here.
What is still to be done is to perform 
the summation of the factorial series of \eq\nr{soln},
which means truncating the infinite sum at some $s_{\rm max}$. 
Studying the convergence behavior of these sums, one
notices that even in the cases where they do converge down to $x\sim 1$, 
their convergence properties usually strongly decline
with decreasing $x$. This means that in practical computations, where one 
aims at obtaining a maximal number of correct digits for
$U(1)$ with as little CPU time as possible, the optimal strategy is to 
evaluate the integral $U(x)$ with the factorial series
approach at some $x_{\rm max}\gg 1$ and then use the recurrence relation of 
\eq\nr{diffeqn} to obtain the desired result at $x=1$. 
The price to pay is, however, a loss of numerical
accuracy at each `pushdown' ($x\rightarrow x-1$) step
due to possible cancellations, which
makes the use of a very high $x_{\rm max}$ impossible. 
In practice the strategy is to
determine an optimal value for the
ratio $s_{\rm max}/x_{\rm max}$.
To give an example, for the four-loop integrals of Section \ref{se:four} 
we have found that $s_{\rm max}/x_{\rm max}\sim 50$ is a good value,
while we used a range of $s_{\rm max}\sim 1350\dots 2000$.


\section{\label{se:three} Implementation of the algorithm}

As is apparent from the preceding section, there are three main steps 
involved in obtaining the desired numerical coefficients in the 
$\e$\/-expansion of each master integral: 
deriving the difference equations obeyed by each integral,
solving them in terms of factorial series, 
and finally performing the $\e$\/-expansion and numerically 
evaluating the sum of \eq\nr{soln} (truncated at $s_{\rm max}$) 
to the precision needed.
We will briefly address each of them in the following.

For the first step, we slightly generalized the IBP algorithm
we had used for reducing generic 4-loop bubble integrals to master
integrals, which follows the setup given in \cite{Laporta:2001dd}, 
and whose implementation in FORM \cite{Vermaseren:2000nd}
is documented in \cite{Schroder:2002re}.
The main difference is an enlarged representation for the
integrals, keeping track of the line which carries the
extra powers $x$, as well as the fact that there are now
two independent variables ($d$, $x$), requiring factorization
(and inversion) of bivariate polynomials, as opposed to univariate
polynomials in the original version.

Second, staying within FORM for convenience, we implemented routines
that straightforwardly solve the difference equations in terms
of factorial series, along the lines of \cite{Laporta:2001dd}. 
This is done starting with the simplest one-loop master integral,
and working the way up to the most complicated (most lines) 
four-loop integral,
ensuring that at each step, the `simpler' terms constituting the
inhomogeneous parts of the difference equation are already known.
The output are then plain ascii files 
specifying each solution in the form of \eq\nr{soln} as well as
containing recursion relations for the coefficients $a(s)$. 
Note that these first two steps are performed exactly, in $d$ 
dimensions.

Third, once the recursion relations for the coefficients $a(s)$ were known, 
we used a Mathematica program to obtain their numerical values at each 
$s$ to a predefined precision, and to perform the summation of the 
factorial series. 
While this procedure is in principle very straightforward, 
there are some twists that we employed to help reduce the 
running times significantly, most of 
which are probably
quite specific to our use of Mathematica.
To avoid a rapid loss of significant digits in 
solving the recursion steps that relate each $a(s)$ to $a(0)$, especially 
those for the homogeneous coefficients, we first solved the relations 
analytically and only in the end substituted the
numerical value (actually the truncated $\e$\/-expansion) 
of the first non-zero coefficient. 
In fact, we found Mathematica to operate quite efficiently
with operations like multiplication of two truncated power series,
so that we relied heavily on it.
Furthermore, since --- not surprisingly --- the
most time-consuming part in the summation of the series
turned out to be the $\e$\/-expansion of $\Gamma$\/-functions, 
we achieved a notable speed-up by substituting the $\Gamma$\/-functions with
large arguments by suitable products of linear factors times 
$\Gamma$\/-functions of smaller arguments.
Finally, a vital step in avoiding an excessive loss in the
depth of the $\e$\/-expansions when going
from one integral to the next, was to apply the
`Chop' command to remove from the results and coefficients excess 
unphysical poles, whose coefficients were of the
order of, say, $10^{-50}$ or less.


\section{\label{se:four} Numerical results}

Below we list the Laurent expansions in $\e=(3-d)/2$ of the 
1+1+3+13 fully massive vacuum master integrals up to four loops.
We use an intuitive graphical notation, in which 
each line represents a massive scalar propagator, 
while dot on a line means it carries an extra power.
The integral measure we have chosen here is
\ba
\int_p &\equiv& \fr{1}{\Gamma(3/2+\e)}\int\!
\fr{{\rm d}^{3-2\e}p}{\pi^{3/2-\e}}.
\ea
In each case\footnote{With the exception of the 
last two integrals, for which we were at this
time able to produce only the first 6 and 5 $\e$\/-orders, respectively.} we 
provide the first 8 $\e$\/-orders keeping the
accuracy at 50 significant digits for the 1-, 2-, and 3-loop master 
integrals and at 22-25 for the 4-loop ones. 
To obtain more $\e$\/-orders and significant digits is merely 
a matter of additional CPU time.

\newcommand{\nN}{\nn[-3pt]}
\ba
\one \label{J}
&=&{}  -4.0000000000000000000000000000000000000000000000000 
\nN&&{}-16.000000000000000000000000000000000000000000000000 \ep^2
\nn&&{}-64.000000000000000000000000000000000000000000000000 \ep^4
\nn&&{}-256.00000000000000000000000000000000000000000000000 \ep^6
       +\order{\e^{8}}\\
\two \label{S}
&=&{}  +4.0000000000000000000000000000000000000000000000000 \ep^{-1}
\nN&&{}-14.487441729730630111648209847429586185151846775400 
\nn&&{}+41.495035953369978394225958244504121655360756728405 \ep
\nn&&{}-107.49752321579967383991953818365893067117808339742 \ep^2
\nn&&{}+263.49878761720606330238135348797499506915058750280 \ep^3
\nn&&{}-623.49940078392000186832902635721463645559035022216 \ep^4
\nn&&{}+1439.4997026869879573968449524699557874962297882621 \ep^5
\nn&&{}-3263.4998520860644726225542919399943895943491031166 \ep^6
       +\order{\e^7}\\
\threeB
&=&{}  -64.00000000000000000000000000000000000000000000000 \ep^{-1}
\nN&&{}+49.44567822334599921081142309329320142732803439623
\nn&&{}-1981.207736229513534030093683214422278348416661525 \ep
\nn&&{}-235.7077170926718752095474374908098006136204356228 \ep^2
\nn&&{}-63521.71508871044639640714223746384514019533126715 \ep^3
\nn&&{}-33675.11111780076696716334804652776927940758434016 \ep^4
\nn&&{}-2213147.071275511251113640247844877948334091419700 \ep^5
\nn&&{}-1414250.728717593474053272387541196652013773984236 \ep^6
+\order{\e^7}\\
\threeV \label{V}
&=&{}  +32.859770043923503738827172731532536947448547448996
\nN&&{}-365.41238154175547388711920818936800707879030719734 \ep
\nn&&{}+2803.7940402523167047150293858439985472095966118207 \ep^2
\nn&&{}-18727.187392108144301607279844058527418378836943988 \ep^3
\nn&&{}+117794.35873133306139734878960626307962150043480498 \ep^4
\nn&&{}-721386.63300305569920915438185951112611780543107044 \ep^5
\nn&&{}+4366100.1639736899128559563097848872427318803864139 \ep^6
\nn&&{}+26291285.708454833832306242766439811661977583440814 \ep^7
+\order{\e^{8}}\\
\threeM \label{merc}
&=&{} +1.391204885296021941812048136925327740910466706390
\nN&&{}-4.898152455251800666032641168608190942446944333758 \ep
\nn&&{}+12.98842503803858164353982398007130232261458098462 \ep^2
\nn&&{}-30.39637625288207454078370310227949470365033235457 \ep^3
\nn&&{}+66.67957617359017942652215661267829752624475575093 \ep^4
\nn&&{}-140.9974945708845413812214824315460314748605690042 \ep^5
\nn&&{}+291.7287632268179138442199742398614147733926624689 \ep^6
\nn&&{}-595.7006275449402266695675282375932229509102799733 \ep^7
+\order{\e^{8}}\\
\topoXII 
&=&{}  +720.0000000000000000000000 \ep^{-1}
       -52.13034199729620858728708
\nN&&{}+33748.69042965137616701638 \ep
       +10819.60558535024688749473 \ep^2
\nn&&{}+1311729.690542895866693548 \ep^3
       +615270.7589383441011319577 \ep^4
\nn&&{}+48899219.67276170476701364 \ep^5
       +24885879.11003549349511900 \ep^6
       +\order{\e^7}\\
\topoXIIxtra \label{xtraA}
&=&{}  -32.00000000000000000000000 \ep^{-1}
       +21.28521367989184834349148
\nN&&{}-945.4764617862257950102533 \ep
       -500.9879407913869195081538 \ep^2
\nn&&{}-29027.99548541518650323471 \ep^3
       -34796.65982174097113175672 \ep^4
\nn&&{}-993306.5068744076465770453 \ep^5
       -1406349.173668893367086333 \ep^6
       +\order{\e^7}\\
\topoIX 
&=&{}  +8.000000000000000000000000 \ep^{-2}
       -25.94976691892252044659284 \ep^{-1}
\nN&&{}-152.5193565764658289654545
       +2653.873458838396323815566 \ep
\nn&&{}-23471.05910309626447406639 \ep^2
       +169839.2007120049515774452 \ep^3
\nn&&{}-1124117.877397355450165203 \ep^4
       +7116455.837989754857686241 \ep^5
       +\order{\e^6}\\
\topoVIII \label{T}
&=&{}  +78.95683520871486895067593 \ep^{-1}
       -1062.608419332108844057560
\nN&&{}+9340.076804859596283223881 \ep
       -68699.47293187699594375521 \ep^2
\nn&&{}+462145.6926820632806821051 \ep^3
       -2963063.672524354359852913 \ep^4
\nn&&{}+18494675.22629230338091457 \ep^5
       -113673206.9834859509114931 \ep^6
       +\order{\e^7}\\
\topoVI 
&=&{}  +33.05150971425671642138224
       -358.4595946559340238066389 \ep
\nN&&{}+2451.469078369636793421997 \ep^2
       -13564.14170819716549262162 \ep^3
\nn&&{}+66602.55178881628657891800 \ep^4
       -303915.1384697444382333780 \ep^5
\nn&&{}+1323370.670112542076081095 \ep^6
       -5589978.086026239748023404 \ep^7
       +\order{\e^{8}}\\
\topoV 
&=&{}  +27.57584879577521927818358
       -291.4075344540614879796315 \ep
\nN&&{}+1956.162997112043390446958 \ep^2
       -10678.5639091187201818981 \ep^3
\nn&&{}+51925.3888799007705970928 \ep^4
       -235296.36309585614167636 \ep^5
\nn&&{}+1019555.9650538012793966 \ep^6
       -4292011.3101269758990557 \ep^7
       +\order{\e^{8}}\\
\topoIV 
&=&{}  +19.84953756526739935782082
       -200.9768306606422068619864 \ep
\nN&&{}+1308.883448000100198800887 \ep^2
       -6990.22562100063537185149 \ep^3
\nn&&{}+33456.8326902483214417013 \ep^4
       -149903.697032731221510018 \ep^5
\nn&&{}+644404.61801211590204150 \ep^6
       -2697912.0878890801856234 \ep^7
       +\order{\e^8}\\
\topoIVxtra \label{xtraB}
&=&{}  +3.141336279450209755917806
       -19.78740273338730374386071 \ep
\nN&&{}+83.81604328128850410126511 \ep^2
       -295.3496021971085625102731 \ep^3
\nn&&{}+934.2247995435558122394582 \ep^4
       -2751.31852347627462886909 \ep^5
\nn&&{}+7700.18972963585089750348 \ep^6
       -20740.9769474365145116212 \ep^7
       +\order{\e^8}\\
\topoIII 
&=&{}  +2.012584635078182771827701
       -10.76814227797251921324485 \ep
\nN&&{}+39.40636857271936487899035 \ep^2
       -121.0015646826735646109733 \ep^3
\nn&&{}+335.6942965583773421544251 \ep^4
       -872.009773755552224781319 \ep^5
\nn&&{}+2163.88707221986880315576 \ep^6
       -5193.51249188593850483093 \ep^7
       +\order{\e^8}\\
\topoII 
&=&{}  +1.27227054184989419939788
       -5.67991293994853579036683 \ep
\nN&&{}+17.6797238948173732343788 \ep^2
       -46.5721846649543261864019 \ep^3
\nn&&{}+111.658522176214385363568 \ep^4
       -252.46396390100217743236 \ep^5
\nn&&{}+549.30166596161426941705 \ep^6
       -1164.5120588971521623546 \ep^7
       +\order{\e^8}\\
\topoIIxtra \label{xtraC}
&=&{} +0.297790726683752651865168
      -0.709896385699143430126726 \ep
\nN&&{}+1.40535549472683132370135 \ep^2
       -2.45721908509256673440117 \ep^3
\nn&&{}+4.00998036005764459707090 \ep^4
       -6.2518071963546459390185 \ep^5
\nn&&{}+9.4402506572040685160665 \ep^6
       -13.924465979877416801887 \ep^7
       +\order{\e^8}\\
\topoI 
&=&{}  +0.233923932580303206470057
       -0.48523164074102176840584 \ep
\nN&&{}+0.88555744401503729577888 \ep^2
        -1.438019871368410241810 \ep^3
\nn&&{} +2.198725350440790755608 \ep^4
        -3.231974794381719679729 \ep^5
        +\order{\e^6}\\
\topoEX \label{X9}
&=&{}  +0.195906401341238799905792
       -0.37006152907989745845214 \ep
\nN&&{}+0.65228273818146302130509 \ep^2
        -1.029288152514143871118 \ep^3
\nn&&{} +1.542484509438506710808 \ep^4
        +\order{\e^5} 
\ea

We have performed various checks in order to
test the correctness of our recursion relations 
as well as to verify the number of 
exact digits contained in our results \eqs\nr{J}-\nr{X9}. 
The first task we have completed by exploiting the fact that the 
recursion relations are not specific to $d=3-2\e$, but can
easily be applied to any dimension, such as $d=4-2\e$. 
We have successfully verified the results of Ref.
\cite{Laporta:2002pg} to somewhat lower accuracy and depth in $\e$. 
Note that our choice of a basis for 4-loop master integrals 
differs slightly from 
the one made in \cite{Laporta:2002pg}. 
The relations needed 
for a basis transformation are listed in \cite{Schroder:2002re}.
An immediate advantage in the light of difference equations
is that with our choice, the above results \eqs\nr{xtraA},\nr{xtraB} and 
\nr{xtraC} follow `for free' from their counterparts without dots.

The accuracy of our three-dimensional
results we have on the other hand examined in three independent ways:
\begin{itemize}
\item by comparing the numerical results to existing analytic calculations;
they can be found in 
\cite{Rajantie:1996cw} (divergent and constant parts of \eqs\nr{S}-\nr{V}), 
\cite{Broadhurst:1998ke} (leading term of \eq\nr{merc}), 
\cite{Braaten:1995cm} (divergence of \eq\nr{T})
and \cite{avgradu,Kajantie:2003ax} (all divergences and some constant parts
of 4-loop integrals, as well as some ${\cal O}(\e)$ terms of lower-loop 
cases).
\item by comparing the results obtained by raising topologically 
inequivalent lines 
to the power $x$,
\item by analyzing the convergence properties of the factorial series, 
i.e.~by checking the stability of our results with respect to
varying $s_{\rm max}$.
\end{itemize}
The first method is of course exact, but is only available for 
a few low (in $\e$) orders for approximately 
half of the integrals considered. 
The second one, on the other hand, has the advantage of covering all 
the different powers of $\e$, but is inapplicable for those 
integrals, in which all propagators are equivalent (e.g. the
basketball-topology). 
The third method is then the most widely applicable 
one, but has the downside of providing no
evidence for the correctness of our results, rather giving only the number 
of digits stable in the variation of the cut-off of the
factorial series. 
For the integral of \eq\nr{X9} only the 
last method is available, but in addition we
have verified the leading term in the result to 3 digits using a 
Monte Carlo integration of an 8-dimensional
integral representation derived for this integral in Ref. \cite{avgradu}.

One might be concerned about the rapid growth with increasing 
$\e$\/-orders of most of the coefficients. 
This is, as was pointed out in \cite{Laporta:2002pg}, caused
by poles that the integrals (seen as functions of $d$) develop near 
$d=3$, e.g. at $d=7/2, 4$, etc. 
It is to be expected that factoring out the first few of these nearby 
poles in each case will improve the apparent convergence in $\e$
considerably.

In principle, having a method at hand that is capable of
generating coefficients to very high accuracy, even to a couple
of hundred digits, one could now use the algorithm PSLQ \cite{pslq}
combined with an educated guess of the number content of some 
of the yet-unknown constant terms, in order to search for
analytic representations of the numerical results. These could
then in turn be used as an inspiration to find useful
transformations of the integral representation of the 
original integral, which might allow for a fully analytic solution
in those cases where it could not yet be achieved.
We have not made any attempts in that direction, since 
the numerical accuracy of the results \eqs\nr{J}-\nr{X9}
should be sufficient for all practical purposes.


\section*{Acknowledgments}

This research was supported in part
by the DOE, under Cooperative Agreement no.~DF-FC02-94ER40818,
and by the Academy of Finland, Contract no.~77744.
A.V. was also supported by the Foundation of Magnus Ehrnrooth.
Y.S. would like to thank the Department of Physics, Helsinki, for hospitality.
A.V. would like to thank the CTP, Cambridge, for hospitality.
We are grateful to Ari Hietanen for helping us provide an
independent check of the leading term of \eq\nr{X9}.



\end{document}